\documentclass{elsart}

\usepackage{amssymb}
\usepackage{amsmath}
\journal{Physics Letters B}

\begin{document}

\begin{frontmatter}

\title{The Area Quantum and Snyder Space}
\author[ICN]{Juan~M.~Romero},
\ead{sanpedro@nucleares.unam.mx}
\author[DMAS]{Adolfo~Zamora\corauthref{cor}}
\corauth[cor]{Corresponding author.}
\ead{zamora@correo.cua.uam.mx}

\address[ICN]{Instituto de Ciencias Nucleares,
Universidad Nacional Aut\'onoma de M\'exico,\\
Apartado Postal 70-543, M\'exico 04510 
DF, M\'exico}
\address[DMAS]{Departamento de Matem\'aticas 
Aplicadas y Sistemas, Universidad Aut\'onoma\\
Metropolitana -- Cuajimalpa, M\'exico 01120 DF, M\'exico}

\begin{abstract}
We show that in the Snyder space the area of the disc and of the sphere can 
be quantized. It is also shown that the area spectrum of the sphere can be 
related to the Bekenstein conjecture for the area spectrum of a black hole 
horizon.
\end{abstract}

\end{frontmatter}

\section{Introduction}
\label{s:Intro}
One of the fundamental problems in physics is the quantization of gravity. A 
satisfactory solution to it has not been given yet. Nevertheless, there are 
signs that the theory must imply discretization of some geometric quantities. 
For instance, it can be shown from heuristic arguments that the area of the
event horizon of a black hole must have a discrete spectrum. The proof follows
from Ehrenfest principle \cite{ehrenfest:gnus}, which states that to every 
classical adiabatic-invariant there corresponds a discrete spectrum in its 
quantum version. Thus, as this area is proportional to the entropy of the 
black hole, which is a classical adiabatic-invariant, Bekenstein proposed that 
it must have a discrete spectrum in the quantum level of the form 
\cite{Bekenstein:gnus,Bekenstein1:gnus}
\begin{equation}
A_{n}=4\pi r^{2}\approx \gamma l_{p}^{2}n, \quad n=1,2,\dots 
\label{eq:hoyo}
\end{equation}
with $\gamma$ being a proportionality constant and $l_{p}=\sqrt{\hbar G/c^{3}}$
the Planck length. It can be seen that (\ref{eq:hoyo}) is essentially the 
area of a sphere of radius $r$.\\

It is expected that the correct version of quantum gravity be consistent 
with proposal (\ref{eq:hoyo}). That implies a space-time consistent with 
discretization of some geometric quantities in the level of quantum gravity.
In this sense, G. 't Hooft showed that in quantum gravity in $(2+1)$ 
dimensions a discrete space-time is obtained in an effective fashion 
\cite{hooft:gnus}. This result suggests that in other dimensions quantum
gravity implies discrete spaces. It is worth pointing out that the discrete
space obtained by G. 't Hooft is the so-called Snyder space, which is
interesting because it is discrete, noncommutative and compatible with the 
Lorentz symmetry \cite{snyder:gnus}. In fact, based on it other 
Lorentz-invariant discrete spaces can be constructed 
\cite{yang:gnus,tanaka:gnus,nos2:gnus}. A proposal to quantize gravity 
based on discrete spaces can be seen in \cite{sorkin:gnus}.\\

In this letter we show that, starting from the Snyder space, the area of the 
disc and of the sphere can be quantized. We also show that the area spectrum 
of the sphere can be related to the Bekenstein conjecture (\ref{eq:hoyo}).\\

The manuscript is organized as follows: in Section \ref{s:TSS}
we study some properties of the Snyder space. We then use these properties
to calculate the area quantum: of the disc in Section \ref{s:AD} and of the 
sphere in Section \ref{s:AS}. Finally, in Section \ref{s:Summ} we summarize 
our results.\\

\section{The Snyder Space}
\label{s:TSS}
We shall express the Snyder space in terms of the 
$\zeta^{A}=(\zeta^{0},\zeta^{1},\cdots,\zeta^{D},\zeta^{D+1})$ variables
which are considered in a flat metric with signature 
$sig(\eta_{AB})=(1,-1,\cdots,-1,-1)$. In this fashion, the actual coordinates 
of the Snyder space \cite{snyder:gnus} and their commutation rules are given 
by
\begin{eqnarray}
\hat x^{\mu}&=&-ia\left(\zeta^{D+1}\frac{\partial}{\partial \zeta_{\mu}}
- \zeta^{\mu}\frac{\partial}{\partial \zeta_{D+1}}\right),\\
\left[\hat x^{\mu},\hat x^{\nu}\right]&=& ia \hat l^{\mu\nu},
\label{eq:XmuXnu}\\
\hat l^{\mu\nu} &=& -ia\left(\zeta^{\nu}
\frac{\partial }{\partial \zeta_{\mu}}-\zeta^{\mu}
\frac{\partial }{\partial \zeta_{\nu}}\right),\\
& & \mu,\nu=0,1,\cdots, D.
\end{eqnarray}
where $a$ is a constant with units of length and $\zeta^{D+1}$ is a parameter 
invariant under the $SO(D,1)$ group \cite{snyder:gnus}. In particular, for the
$D=3$ case, the commutation rules of the spatial variables can be represented
by the matrix
\begin{equation}
C_{ij}=
\left(
\begin{array}{ccc}
-\hat y & \hat x & \hat l^{12} \\
-\hat z & \hat y & \hat l^{23} \\
-\hat x & \hat z & \hat l^{31}
\label{eq:Cij}
\end{array}
\right),\qquad 
\hat x=\hat x^{1},\quad \hat y=\hat x^{2}, \quad \hat z=\hat x^{3}.
\end{equation}
Notice that the rows of this matrix 
satisfy the commutation rules
\begin{equation}
\left[ C_{li}, C_{lj} \right]= ia\epsilon_{ijk} C_{lk}.
\label{eq:CliClj}
\end{equation}
That is, for every pair of spatial variables and their commutator the $SO(3)$ 
algebra holds, i.e. a noncommutative space of the type of the fuzzy sphere 
\cite{madore:gnus}.\\

It can be shown that $\hat l^{\mu\nu}$ is the generator of Lorentz 
transformations in the Snyder space \cite{nos2:gnus}. In fact, for the 
spatial part of the commutation rules, (\ref{eq:XmuXnu}) can be written as
\begin{equation}
\left[\hat x_{i},\hat x_{j}\right]=
ia\epsilon_{ijk}\hat l_{k}, \quad \hat l_{i}=-ia\epsilon_{ijk}\zeta^{j}
\frac{\partial }{\partial \zeta^{k}},
\end{equation}
where $\hat l_{i}$ is the generator of rotations which satisfies
\begin{eqnarray}
\left[\hat l_{i},\hat l_{j}\right]&=&ia\epsilon_{ijk}\hat l_{k}, \\
\left[\hat x_{i},\hat l_{j}\right]&=&ia\epsilon_{ijk}\hat x_{k}.
\end{eqnarray}
By using $\hat l_{i}$ and $\hat x_{j}$, one may define the operators
\begin{eqnarray}
\hat M_{i}&=& \frac{1}{2}\left(\hat l_{i}+\hat x_{j}\right),
\label{eq:Mi}\\
\hat N_{i}&=&\frac{1}{2}\left(\hat l_{i}-\hat x_{j}\right),
\label{eq:Ni}
\end{eqnarray}
which satisfy
\begin{eqnarray}
\left[\hat M_{i},\hat M_{j}\right]&=&ia\epsilon_{ijk}\hat M_{k}, \\
\left[\hat N_{i},\hat N_{j}\right]&=&ia\epsilon_{ijk}\hat N_{k},\\
\left[\hat N_{i},\hat M_{j}\right]&=&0.
\end{eqnarray}
That is, the algebra of $SO(3)\times SO(3)$. In this sense the spatial
sector of the Snyder space can be regarded as a combination of two
fuzzy spheres.\\

\section{Area Quantum of the Disc}
\label{s:AD}
To obtain this quantum we use the spatial part of the commutation rules 
(\ref{eq:XmuXnu}), which are equivalent to (\ref{eq:Cij}). In 
particular, for the $x$-$y$ plane we define the operators 
\begin{eqnarray}
\hat L^{2}&=&\hat x^{2}+\hat y^{2}+\left(\hat l^{12}\right)^{2}, \\
\hat {\cal A}&=&\pi \left(
\hat L^{2}- \left(\hat l^{12}\right)^{2}\right)=
\pi \left(\hat x^{2}+\hat y^{2}\right),
\end{eqnarray}
where $\hat{\cal A}$ can be identified with the area operator of a disc.
By using the fact that the rows of matrix (\ref{eq:Cij}) satisfy the
$SO(3)$ algebra, Eq. (\ref{eq:CliClj}), one finds the spectra
\begin{eqnarray}
\hat L^{2}|nm\rangle &=&L^{2}_{nm}|nm\rangle
=  a^{2}n(n+1)|nm\rangle, \\
\hat {\cal A}|nm\rangle &=&{\cal A}_{nm} |nm\rangle
=\pi a^{2} \left[n(n+1)-m^{2}\right]|nm\rangle,\qquad \\
 & \hbox{for} & n=0,1,2,\cdots\ \ \hbox{and}\ \ 
-n\leq m\leq n, \label{eq:aq}
\end{eqnarray}
so that the area of the disc, $A_{nm}$, is discrete.\\

It can be shown that the uncertainty principle
\begin{equation}
\Delta x\Delta y\geq \frac{1}{2}|\left[\hat x,\hat y\right]|,
\end{equation}
for this case becomes
\begin{equation}
\frac{a^{2}}{2}\left[n(n+1)-m^{2}\right]\geq \frac{a^{2}}{2}|m|.
\end{equation}
Notice that the equality holds when $m=\pm n$, which is the case of the 
coherent states. Thus, for the coherent states the area of the disc has
spectrum
\begin{equation}
{\cal A}_{nn}=\pi a^{2}n.
\end{equation}
We remark that, for a given $n$, the uniformly spaced spectrum holds when 
$|m|$ reaches its largest value, which occurs when the area is a minimum.\\

All the previous results hold equally for the $x$-$z$ and $y$-$z$ planes. 
Another noncommutative space that allows to quantize the area of the disc 
can be seen in \cite{nos1:gnus}.

\section{Area Quantum of the Sphere}
\label{s:AS}
To get this quantum let us recall that in Minkowski space
\begin{equation}
ds^{2}=c^{2}(dt)^{2}-(dr)^{2}-r^{2}d^{2}\Omega
\end{equation}
whereas in Schwarzschild space
\begin{equation}
ds^{2}=\left(1-\frac{2MG}{c^{2}r}\right)c^{2}(dt)^{2}
-\left(1-\frac{2MG}{c^{2}r}\right)^{-1}(dr)^{2}-r^{2}d^{2}\Omega,
\end{equation}
so that the area of the sphere in each case is
\begin{equation}
A=4\pi r^{2}.
\end{equation}
Therefore, in principle, the spectrum of $A$ must be the same in either the 
Minkowski or Schwarzschild space. Then, as quantizing $A$ is equivalent to 
quantizing $r^{2}=x^{2}+y^{2}+z^{2}$, we use Eqs. (\ref{eq:Mi}--\ref{eq:Ni})
to find
\begin{eqnarray}
\hat r^{2}&=&\hat x^{2}+\hat y^{2}+\hat z^{2}=(\hat M-\hat N)^{2}=
\hat M^{2}+\hat N^{2}-2\hat N\cdot \hat M\\
\hat l^{2}&=&(\hat M+\hat N)^{2}=\hat M^{2}+\hat N^{2}+2\hat N\cdot \hat M,
\end{eqnarray}
which can be combined to get
\begin{equation}
\hat A=4\pi\hat r^{2}=4\pi\left(2\left(\hat M^{2}+\hat N^{2}\right)-\hat l^{2}
\right).
\end{equation}
By using the addition rule of two angular momenta one gets to the spectrum 
\cite{landau:gnus}
\begin{equation}
A_{l_{1}l_{2}}=
4\pi a^{2}\Bigg(2\bigg(l_{1}(l_{1}+1) +l_{2}(l_{2}+1)\bigg)
-l(l+1)\Bigg),
\end{equation}
where $l_{1},l_{2}=0,1,2,3,\cdots$ and $|l_{1}-l_{2}| \leq l \leq l_{1}+l_{2}$.
Notice that in the case $l_{1}=l_{2}=n$ and $l=2l_{1}=2n$, one finds the 
spectrum 
\begin{equation}
A_{n}= 8\pi a^{2}n, \label{eq:abek}
\end{equation}
which coincides with Bekenstein conjecture (\ref{eq:hoyo}). Comparison between
spectra (\ref{eq:abek}) and (\ref{eq:hoyo}) suggests that the constant $a$ 
must be of the order of the Planck length.\\

Analogously to the disc, for $l_{1}=l_{2}$ the uniformly spaced spectrum is 
obtained when $l$ is maximum; that is when the area is a minimum.\\

It should be pointed out that, starting from the Snyder space, Yang constructed
a discrete space-time which, apart from being compatible with the Lorentz
symmetry, is also compatible with the Poincare Symmetry \cite{yang:gnus}. The
commutation rules in the spatial coordinates of the Yang space coincide with
those of the Snyder space and so the results here presented are also valid
for the Yang space.\\

The fact that the Snyder space is compatible with Bekenstein conjecture
(\ref{eq:hoyo}) reinforces the idea that in some limit quantum gravity in
$(3+1)$ dimensions implies the Snyder space. Other works relating quantum
gravity and noncommutative spaces can be seen in \cite{freidel:gnus}.\\

\section{Summary}
\label{s:Summ}
It was shown that the spatial sector of the Snyder space can be regarded as an
array of fuzzy spheres. In addition, the area quantum of the disc and of the 
sphere were obtained. It was also shown that an uniformly spaced spectrum may
be obtained for the sphere which coincides with the Bekenstein conjecture for 
the area of a black hole horizon. This last result reinforces the idea that in
some limit quantum gravity in $(3+1)$ dimensions may imply the Snyder space.\\

\section*{Acknowledgments}
\label{s:ack} 
JMR wishes to thank the Instituto de Ciencias Nucleares for kind hospitality.

\end{document}